\documentclass[12p]{article}
\usepackage{latexsym}
\usepackage{color}
\usepackage{amsmath}
\usepackage{epsfig}
\usepackage{hyperref}
 \usepackage{dcolumn}
   \usepackage{threeparttable}

\newcommand{\ortala}[1]{\begin{center}#1\end{center}}

\newcommand{\ket}[1]{\left|#1\right\rangle}

\newcommand{\sand}[3]{\left\langle #1\left|#2\right|#3\right\rangle}
\newcommand{\sandd}[1]{\left\langle #1\right\rangle}

\newcommand{\summ}[3]{{{\underset{#1 }{\overset{#2}{\displaystyle\sum}}}#3}}
\newcommand{\prodd}[3]{{{\underset{#1
}{\overset{#2}{\displaystyle\prod}}}#3}}

\newcommand{\re}[1]{(\ref{#1})}

\newcommand{\eq}[2]{\begin{equation}\label{#1}  #2\end{equation}}

\newcommand{\paran}[1]{\left(#1\right)}

\newcommand{\sch}[1]{Schrodinger}
\newcommand{\ktur}[2]{\frac{\partial #1}{\partial #2}}

\newcommand{\komb}[2]{\paran{\begin{array}{c} #1 \\ #2 \end{array}}}

\usepackage{amssymb}
\usepackage{latexsym}
\begin{document}

\ortala{\large\textbf{Effective field theory in larger clusters - Ising Model}}

\ortala{\textbf{\"Umit Ak\i nc\i \footnote{\textbf{umit.akinci@deu.edu.tr}}}}

\ortala{\textit{Department of Physics, Dokuz Eyl\"ul University,
TR-35160 Izmir, Turkey}}

\section{Abstract}

General formulation for the effective field theory with differential operator
technique and the decoupling approximation with larger finite clusters (namely EFT-$N$ formulation) has been derived,
for S-1/2 bulk systems. The effect of the enlarging this finite cluster on the results in the critical temperatures
and thermodynamic properties have been investigated in detail. Beside the improvement on the critical temperatures,
the necessity of using larger clusters, especially in nano materials have been discussed. With the derived formulation, application
on the effective field and mean field renormalization group techniques also have been performed.

\section{Introduction}\label{introduction}

Cooperative phenomena in magnetic systems are often investigated within
some approximation methods in statistical physics. There are still a
few exact results in the literature \cite{ref1}, since the partition
function is not tractable in most of the systems. The most known example of
this situation is that there is still no exact result for the most basic
model of magnetic systems, namely  Ising model \cite{ref2} in
three dimensions, although exact result  for two dimensional system
has been presented in 1944 \cite{ref3}. There are numerous
approximation and simulation methods for these systems. Each of
these  methods have their own advantages as well as disadvantages.
A class of these approximation methods is called effective field
theories (EFT) \cite{ref4}. Recent developments in these
formulations, especially in correlated effective theories can be
found in Ref. \cite{ref5}.

Early attempts to solve Ising model yields mean field theories
(MFT), which reduce the many particle Hamiltonian into one particle, 
with replacing the spin operators in the Hamiltonian with their
thermal (or ensemble) averages. This means that neglecting all
self-spin and multi-spin correlations in the system. After than,
by handling the self-spin correlations, EFT formulations have been
constructed. First successful variants of these approximations are
Oguchi approximation \cite{ref6} and Bethe-Peierls approximation
(BPA) \cite{ref7,ref8}. After than, many variants of the EFT
constructed with their own advantages, disadvantages and own
limitations \cite{ref5}.

Most of the EFT formulations start by constructing a finite
cluster within the system. Interactions between the spins which are
located in this cluster are written exactly as much as possible and the
coupling of this cluster with the outside of it is written
approximately. The problem arises when we work with finite
clusters which represent the whole system. Let us call the
spins located in the chosen cluster as inner spins, spins located at
the borders of the chosen finite cluster as border spins and all
other spins as outer spins, i.e. outer spin is any spin which is
outside of the chosen cluster. The interactions between the inner
spins with other inner spins or border spins  can be calculated with a
given Hamiltonian of the system. The problem comes from the
interactions of the border spins with their nearest neighbor outer
spins. These interactions have to take into account an
approximate way. In a typical MFT for these systems, this
approximation can be made via replacing all these nearest neighbor
outer spin operators with their  thermal (or ensemble) average.
Although in the spirit of the mean field approximation, reducing the
many particle system to the one particle system, we may call
aforementioned approximation for N-spin cluster as MFT-$N$.

On the other hand EFT can include the self-spin correlations in the
formulation. Then, it is superior to the MFT. One class of the EFT
for the Ising model start with a single-site kinematic relations,
which gives the magnetization of the system, such as Callen identity
\cite{ref9}   or Suzuki identity \cite{ref10}. Although these types of
identities are exact, since they are in a transcendental form,
calculation with   these identities requires some approximations.
Most widely used method here is differential operator technique
\cite{ref11}. Neglecting the multi-spin correlations within this method,
namely using decoupling approximation (DA) \cite{ref12}, it gives the
results of the Zernike approximation \cite{ref13}.  In order to
reduce that transcendental function  given in the Callen identity to
a polynomial form, there are also combinatorial techniques
\cite{ref14,ref15},  integral operator technique \cite{ref16} and
probability distribution technique \cite{ref17}.

On the other hand, larger clusters for obtaining critical properties
of the Ising model for several lattices have been used. For
instance,  $2$-spin cluster (EFT-$2$) \cite{ref18}   and $4$-spin
cluster (EFT-$4$)  \cite{ref19} have been successfully applied to the
Ising systems. But, to the best of our knowledge, there is no general
formulation for EFT-$N$ given. Besides, working with larger clusters
is important for obtaining the critical temperature of the system
within the renormalization group technique, which are within the
mean field renormalization group (MFRG) \cite{ref20}  and effective
field renormalization group (EFRG) \cite{ref21,ref22}  techniques
for the Ising model. Using larger clusters give more closer critical
temperatures to the exact ones. For instance clusters up to number
of 6 spins for the honeycomb lattice, number of 9 spins for the
square lattice and 8 spins for the simple cubic lattice have been
used within the EFRG and more accurate critical temperatures has
been obtained \cite{ref23}.

As seen in this brief literature, working with larger clusters are
important for obtaining more accurate results for the critical and
thermodynamical properties of the Ising model. Since enlarging
the cluster comes with some computational cost, it is important to
answer the question: how large is it enough? Besides, as discussed in
Ref. \cite{ref24}, for the Heisenberg model in nano materials, it is
not an arbitrary choice to use larger clusters, but it is necessity
in some of the systems. This point will be discussed again in later
sections. In the light of these points,  the aim of this work is to
construct a general EFT-$N$ formulation for arbitrary lattice and
compare the results of the solutions in different sized clusters and
exact ones. For this aim, the paper is organized as follows: In Sec.
\ref{formulation} we briefly present the model and  formulation. The
results and discussions are presented in Sec. \ref{results}, and
finally Sec. \ref{conclusion} contains our conclusions.


\section{Model and Formulation}\label{formulation}

We start with  a standard spin-1/2 Ising Hamiltonian with external
magnetic field, \eq{denk1}{
\mathcal{H}=-J\summ{<i,j>}{}{S_iS_j}-H\summ{i}{}{S_i}, } where $S_i$
denotes the $z$ component of the Pauli spin operator at a site $i$,
$J$stands for the exchange interactions between the nearest neighbor
spins and $H$ is the longitudinal magnetic field at any site. The
first summation is carried over the nearest neighbors of the
lattice, while the second one is over all the lattice sites.

In a typical EFT-$N$ approximation, we start with constructing
the $N$-spin cluster and writing $N$-spin cluster Hamiltonian as
\eq{denk2}{
\mathcal{H}^{(N)}=-J\summ{<i,j>}{}{S_iS_j}-\summ{i=1}{N}{h_iS_i}, }
where the first summation is over the nearest neighbor pairs of the
inner and border spins, while the second  summation is over all the
inner and border spins. Here $h_i$ is the local field on a site $i$
and it denotes all the interactions between the  border spin at a
site $i$ and the outer nearest neighbor spins of it  and magnetic
field at a site $i$. We note here that, not all of the inner spins
are the border spins. In this case in this summation some of the
$h_i$ terms may be zero (for the inner spins that are not border
spins at the same time). The term $h_i$ may be called as mean field
or effective field which depends on how we handle it.  Let the site $i$
has the number of $\delta_i$ nearest neighbor outer spins, then
$h_i$ can be written as \eq{denk3}{
h_i=J\summ{k=1}{\delta_i}{S_i^{(k)}}+H, } where $S_i^{(k)}$ denotes
the $k^{\mathrm{th}}$ outer nearest neighbor of the spin $i$ and  $\delta_i$
stands for the number of nearest neighbor outer spins of the spin
$i$. Then   we try to calculate the thermal average of  the quantity
$S_i$ via
\eq{denk4}{ \sandd{S_i} =\sandd{\frac{Tr_N S_i
\exp{\paran{-\beta\mathcal{H}^{(N)}}}}{Tr_N \exp{\paran{-\beta
\mathcal{H}^{(N)}}}}} .} 
In Eq.  \re{denk4}  $Tr_N$ stands for the
partial trace over all the lattice sites which are belonging to the
chosen cluster, $\beta=1/(k_BT)$ where $k_B$ is the Boltzmann
constant, and $T$ is the temperature. Replacing  $S_i$ with some
other quantity related to the system will give the thermal
expectation value of that quantity.  Calculation with Eq. \re{denk4}
requires the matrix representation of the related operators in
chosen basis set, which can be denoted by $\{\psi_i\}$, where
$i=1,2,\ldots 2^N$. Each of the element of this basis set can be
represented by  $\ket{s_1s_2\ldots s_N}$, where $s_k=\pm 1,
(k=1,2,\ldots ,N) $ is just one-spin eigenvalues of the $z$ component
of the spin-1/2 Pauli spin operator. In this representation of the
basis set, operators in the $N$-spin cluster act on a base via

\eq{denk5}{
\begin{array}{lcl}
S_i\ket{\ldots s_i \ldots }&=&s_i\ket{\ldots s_i \ldots }\\
S_iS_j\ket{\ldots s_i \ldots s_j \ldots }&=&s_is_j\ket{\ldots s_i \ldots s_j \ldots }.
\end{array}
}

It is trivial from Eq. \re{denk5} that, matrix representation of the
Eq. \re{denk2} is diagonal, then just calculating the
$\sand{\psi_i}{ -\beta \mathcal{H}^{(N)}}{\psi_i}$ then exponentiate
it is enough for the calculating of Eq. \re{denk4}. Let the diagonal
elements of the matrix representation of $\mathcal{H}^{(N)}$ be
\eq{denk6}{ r_i=\sand{\psi_i}{\mathcal{H}^{(N)}}{\psi_i},} and the
diagonal elements of the matrix representation of the $S_k$ in the
same basis set be \eq{denk7}{ t_i^{(k)}=\sand{\psi_i}{S_k}{\psi_i}
.} Eq. \re{denk4} can be written by using Eqs. \re{denk6} and
\re{denk7} as \eq{denk8}{
m_k=\sandd{S_k}=\sandd{\frac{\summ{i=1}{2^N}{}t_i^{(k)}\exp{\paran{-\beta
r_i}}}{\summ{i=1}{2^N}{}\exp{\paran{-\beta r_i}}}}, k=1,2,\ldots, N
.} The order parameter (i.e. magnetization) of the system can be
defined by \eq{denk9}{ m=\frac{1}{N}\summ{k=1}{N}{m_k }. }

Eq. \re{denk8} can be written in a closed form as \eq{denk10}{
m_k=\sandd{f_k\paran{\beta,J,\{h_i\}}} .} Here, $\{h_i\}$ is stands
for the ordered array of the local fields ($h_1,h_2,\ldots, h_N$) for
the $N$-spin cluster. Thus, the order parameter can be given by
writing Eq. \re{denk10} into Eq. \re{denk9} as

\eq{denk11}{
m=\sandd{F\paran{\beta,J,\{h_i\}}}
} where

\eq{denk12}{
F\paran{\beta,J,\{h_i\}}=\frac{1}{N}\summ{k=1}{N}{}f_k\paran{\beta,J,\{h_i\}}} and

\eq{denk13}{
f_k\paran{\beta,J,\{h_i\}}=\frac{\summ{i=1}{2^N}{}t_i^{(k)}\exp{\paran{-\beta
r_i}}}{\summ{i=1}{2^N}{}\exp{\paran{-\beta r_i}}} } which is nothing but 
the function given in Eq. \re{denk8}.

In literature there are some methods related to evaluation of
the thermal average in Eq. \re{denk11}. Most basic evaluation of the
thermal average is, taking the local fields as \eq{denk14}{
h_i=\delta_i Jm+H } which will give the results of the MFT. It
replaces the outer spin operators  with their thermal (or ensemble)
averages. Note that,  translational invariance property of the
lattice has been used. This means that all sites of the lattice are
equivalent.  With writing Eq. \re{denk14} into Eq. \re{denk11} we
can get the MFT-$N$ equation as \eq{denk15}{
m=F\paran{\beta,J,\{\delta_i J m+H\}} .} Using MFT means 
neglecting the self-spin correlations as well as multi-spin
correlations. We note that, the dependence of the function on the
$\beta$ and $J$ will not be shown in the reminder of the text.

On the other hand, formulations that give better results than the
MFT are presented. One of the class that includes the self spin
correlations in the formulation is EFT. The evolution of Eq.
\re{denk11} is possible in different ways such as differential
operator technique \cite{ref11}, integral operator technique
\cite{ref16} and probability distribution technique \cite{ref17}.

In order to get the explicit form of the order parameter expression
we still have to use some approximations, due to the intractability
of this expression. All approximations produce results within
different accuracy.  For instance,  evaluating Eq. \re{denk11} with
using differential operator technique and DA \cite{ref12} will give
results of Zernike approximation\cite{ref13}. This approximation is
most widely used for these type of systems within the EFT
formulations. Thus, we want to try  using this approximation in
larger clusters. Our strategy will be to start with $1$ and $2$-spin
clusters and then generalize the formulation to the $N$-spin
cluster. We mention that most of the studies in related literature
concerns with $1$ or $2$-spin cluster, although limited works 
using $4$-spin cluster have also been presented, 
such as Ref. \cite{ref19}.

\subsection{$1$-spin cluster}\label{formulation1}

The basis set of the $1$-spin cluster is
$\{\ket{1},\ket{-1}\}$. Calculation of Eq. \re{denk11} by using
this basis set will give \eq{denk16}{ m=\sandd{\tanh{\paran{\beta
h_1  }}}. } With using differential operator technique \cite{ref11},
Eq. \re{denk16} can be written as

\eq{denk17}{ m=\sandd{\exp{\paran{h_1\nabla_1}}}F(x_1)|_{x=0}, }
where $\nabla_1=\ktur{}{x_1}$ is the differential operator and the
function is given by \eq{denk18}{ F(x_1)=\tanh{\paran{\beta x_1}}.
}The effect of the exponential differential operator on an arbitrary
function $f(x_1)$ is defined by \eq{denk19}{
\exp{\paran{a_1\nabla_1}}f(x_1)=f(x_1+a_1), } where $a_1$ is an
arbitrary constant.

By writing Eq. \re{denk3} into Eq. \re{denk17} for $1$-spin
cluster, we can write Eq. \re{denk17}, with the defined operator
\eq{denk20}{ \theta_{j}^{(i,k)}=\exp{\paran{J\nabla_j
S_i^{(k)}}}=\left[\cosh{\paran{J\nabla_j}}+S_i^{(k)}
\sinh{\paran{J\nabla_j}}\right]} as \eq{denk21}{
m=\sandd{\prodd{k=1}{\delta_1}{}\theta_{1}^{(1,k)}}F(x_1)|_{x_1=0}
.} Expansion of Eq.\re{denk21}  contains multi-spin correlations
between the spin $1$ and the nearest neighbors of it. With the help
of the  DA, we can obtain tractable form of this expansion, via
neglecting these multi spin correlations \cite{ref12} \eq{denk22}{
\sandd{S_1^{(1)}S_1^{(2)} \ldots
S_{1}^{(n)}}=\sandd{S_1^{(1)}}\sandd{S_1^{(2)}}\ldots
\sandd{S_{1}^{(n)}} } for $n=3,4,\ldots, \delta_1$. On the other
hand, the translational invariance of the lattice dictates the
equivalence of any two sites in the lattice i.e., \eq{denk23}{
m=\sandd{S_1}=\sandd{S_1^{(1)}}=\sandd{S_1^{(2)}}=\ldots
=\sandd{S_{1}^{(\delta_1)}} .} Using these properties given in Eqs.
\re{denk22} and \re{denk23} in Eq. \re{denk21}, we arrive the
expression for the order parameter as \eq{denk24}{
m=\left[\phi_{1}\right]^{\delta_1}F(x_1)|_{x_1=0} } where
\eq{denk25}{ \phi_{i}=\left[\cosh{\paran{J\nabla_i}}+m
\sinh{\paran{J\nabla_i}}\right] .}

Now, writing hyper trigonometric functions in Eq. \re{denk25} in
terms of the exponentials, then inserting Eq. \re{denk25} into Eq.
\re{denk24} then performing the Binomial expansions, we arrive the
expression of the order parameter as

\eq{denk26}{ m=\summ{n_1=0}{\delta_1}{D_{n_1} m^{n_1}} } where
\eq{denk27}{
D_{n_1}=\summ{r_1=0}{\delta_1-n_1}{}\summ{s_1=0}{n_1}{}E_{r_1s_1}^{(\delta_1,n_1)}F\left[\paran{\delta_1-2r_1-2s_1}J\right]
} and \eq{denk28}{ E_{r_1
s_1}^{(\delta_1,n_1)}=\frac{1}{2^{\delta_1}}\komb{\delta_1}{n_1}\komb{\delta_1-n_1}{r_1}\komb{n_1}{s_1}
\paran{-1}^{s_1}.
}

This is the well known and widely used method, namely EFT with
differential operator technique and DA. This method creates
polynomial form of the expression Eq. \re{denk16} as Eq.
\re{denk26}, as order parameter. As we can see from the Eq.
\re{denk27}, in this process we have to evaluate the function
defined in Eq. \re{denk18} many times at the same point through
running the summations in Eq. \re{denk27}, hence the argument of the
function $\paran{\delta_1-2r_1-2s_1}J$ gets the same value many
times. This point seems not to be  create any problem, since we are
faced with simple function as defined in Eq. \re{denk18} and the
evaluation of the function at the same argument cannot create
significant extra time cost. But when we go to larger clusters we
cannot calculate the analytical form of the function, then we have
to make some matrix operations in order to get the evaluation of the
function at a certain point. This may take some time. For this
reason let us use another form of the order parameter expression.
For this aim let us write Eq. \re{denk25} as

\eq{denk29}{
\phi_{i}=\left[\paran{1+m}\exp{\paran{J\nabla_i}}+ \paran{1-m}\exp{\paran{-J\nabla_i}}\right]
.}

Using this form of the operator in Eq.\re{denk24} with Binomial
expansion will yield an alternative form of the order parameter as
\eq{denk30}{ m=\summ{t_1=-\delta_1}{\prime \delta_1
}{C_{t_1}F\paran{t_1J}} } where $\prime$ denotes the increment of
the dummy indices by $2$ and where \eq{denk31}{
C_{t_1}=\komb{\delta_1}{(\delta_1-t_1)/2}A^{(\delta_1+t_1)/2}B^{(\delta_1-t_1)/2}
} and
\eq{denk32}{
A=\frac{1}{2}\paran{1+m},\quad B=\frac{1}{2}\paran{1-m}.
}

We note that, Eq. \re{denk30} is identical to  Eq.\re{denk26}. The
difference is in their form which means to evaluate the function at a
certain point only once when the summation in Eq. \re{denk30} is
running.

\subsection{$2$-spin cluster}\label{formulation2}

The basis set for the $2$-spin cluster is
$\{\ket{11},\ket{1-1},\ket{-11},\ket{-1-1}\}$. If we evaluate 
Eq. \re{denk9} in this basis set, we arrive the expression of the
order parameter as

\eq{denk33}{ m=\sandd{\frac{\sinh{\left[\beta\paran{h_1+h_2
}\right]}}{\cosh{\left[\beta \paran{h_1+h_2
}\right]}+\exp{\paran{-2\beta J}\cosh{\left[\beta\paran{h_1-h_2
}\right]}}}} } which is nothing but the expression obtained in Ref.
\cite{ref18}.



If we write Eq. \re{denk33}  as in Eq. \re{denk21} we get,
\eq{denk34}{
m=\sandd{\prodd{k=1}{\delta_1}{}\prodd{l=1}{\delta_2}{}\theta_{1}^{(1,k)}\theta_{2}^{(2,l)}}F(x_1,x_2)|_{x_1=0,x_2=0},}
where the function is defined by
\eq{denk35}{ F(x_1,x_2)=\frac{\sinh{\left[\beta\paran{x_1+x_2
}\right]}}{\cosh{\left[\beta \paran{x_1+x_2
}\right]}+\exp{\paran{-2\beta J}\cosh{\left[\beta\paran{x_1-x_2
}\right]}}}.}

By applying the same procedure between Eqs. \re{denk21} and
\re{denk24} to Eq. \re{denk34} we get an expression
\eq{denk36}{
m=\left[\phi_{1}\right]^{\delta_1}\left[\phi_{2}\right]^{\delta_2}F(x_1,x_2)|_{x_1=0,x_2=0}
,} then the expression corresponding to Eq. \re{denk26} in
$2$-spin cluster will be
\eq{denk37}{
m=\summ{n_1=0}{\delta_1}{} \summ{n_2=0}{\delta_2}{} D_{n_1n_2}m^{n_1+n_2}
,} where
\eq{denk38}{
D_{n_1n_2}=\summ{r_1=0}{\delta_1-n_1}{}\summ{s_1=0}{n_1}{}\summ{r_2=0}{\delta_2-n_2}{}\summ{s_2=0}{n_2}{}E_{r_1s_1}^{(\delta_1,n_1)}
E_{r_2s_2}^{(\delta_2,n_2)}
F\left[\paran{\delta_1-2r_1-2s_1}J,\paran{\delta_2-2r_2-2s_2}J\right]
.} The coefficients $E_{r_1s_1}^{(\delta_1,n_1)}$ and
$E_{r_2s_2}^{(\delta_2,n_2)}$  have been defined in Eq. \re{denk28}.
On the other hand, $2$-spin cluster counterpart of Eq.
\re{denk30} can be found within the same procedure as the $1$-spin
cluster and it is given by

\eq{denk39}{ m=\summ{t_1=-\delta_1}{\prime
\delta_1}{}\summ{t_2=-\delta_2}{\prime
\delta_2}{}C_{t_1}C_{t_2}F\paran{t_1J,t_2J}, } in which $\prime$ symbol denotes
the increment of the dummy indices by $2$. The coefficients in Eq.
\re{denk39} have been defined in Eq. \re{denk31}.

\subsection{$N$-spin cluster}\label{formulationN}

For the $N$-spin cluster, the magnetization expressions are given in
Eq. \re{denk11} in a closed form. $N$-spin cluster is constructed in
such a way that is, the total number of inner and  border spins are to
be $N$. The spin at a site $i$, $S_i$ has the number of $\delta_i$
outer spins as its nearest neighbors.

As in $1$-spin cluster (Eq. \re{denk21}) or $2$-spin cluster (Eq.
\re{denk34}), here we can write the magnetization as
\eq{denk40}{
m=\sandd{\prodd{k_1=1}{\delta_1}{}\prodd{k_2=1}{\delta_2}{}\ldots\prodd{k_N=1}{\delta_N}{}
\theta_{1}^{(1,k_1)}\theta_{2}^{(2,k_2)}\ldots
\theta_{N}^{(N,k_N)}}F(\{x_i\})|_{\{x_i=0\}},} where $\{x_i\}$
stands for the ordered array   $x_1,x_2,\ldots, x_N$ for the
$N$-spin cluster. The function $F(\{x_i\})$ is nothing but just the 
replacement of all  $h_i$ terms by  $x_i$, in Eq. \re{denk12}. We
note that, expression given by Eq. \re{denk40} is valid for the
lattices that any inner and border spin has no common outer
neighbors. This means that this form of the formulation cannot give
correct results for some certain lattices such as Kagome lattice.

After expanding Eq. \re{denk40} and applying the DA,  we get an
expression for the order parameter as

\eq{denk41}{
m=\prodd{k=1}{N}{}\left[\paran{\phi_{k}}^{\delta_k}\right]F(\{x_i\})|_{\{x_i\}=0}
,} then the expression corresponding to Eq. \re{denk37} for
$N$-spin cluster will be
\eq{denk42}{ m=\summ{n_1=0}{\delta_1}{} \summ{n_2=0}{\delta_2}{}
\ldots \summ{n_N=0}{\delta_N}{} D_{\{n_i\}}m^{n_1+n_2+\ldots + n_N}
,}  where $\{n_i\}$  stands for the ordered array   $n_1,n_2,\ldots,
n_N$ for the $N$-spin cluster. The coefficient is just the
generalization of the coefficient given in Eq. \re{denk38} for
$2$-spin cluster to the $N$-spin cluster and it is given by
\eq{denk43}{ D_{\{n_i\}}=\summ{\{r_i=0\}}{\{\delta_i-n_i\}
}{}\summ{\{s_i=0\}}{\{n_i\}}{}\left[\prodd{k=1}{N}{}E_{r_ks_k}^{(\delta_k,n_k)}\right]
F\paran{\{\paran{\delta_i-2r_i-2s_i}J\}} .} Here, number of $2N$
summations present, which are running from $r_i=0$ to $\delta_i-n_i$ and
$s_i=0$ to $n_i$, where $i=1,2,\ldots, N$. Also the term
$\paran{\delta_i-2r_i-2s_i}J$ represents the  $i^{\mathrm{th}}$ argument of
the function, where $i=1,2,\ldots, N$. The coefficients
$E_{r_ks_k}^{(\delta_k,n_k)}$ in Eq. \re{denk43} are given as in Eq.
\re{denk28}.

By using a similar procedure for obtaining Eq. \re{denk39} from Eq.
\re{denk36}, we can get from Eq. \re{denk41} \eq{denk44}{
m=\summ{t_1=-\delta_1}{\prime \delta_1}{}\summ{t_2=-\delta_2}{\prime
\delta_2}{}\ldots \summ{t_N=-\delta_N}{\prime \delta_N}{}
\left[\prodd{k=1}{N}{}C_{t_k}\right]F\paran{\{t_k J\}}, } where again
$\prime$ denotes the increment of the dummy indices by $2$. The
coefficients in Eq. \re{denk44} have been defined in Eq.
\re{denk31}.

Thus, we can calculate the order parameter of the system in EFT-$N$
approximation from Eq. \re{denk42} or the equivalent form of it
given in Eq. \re{denk44}, while within the MFA-$N$ approximation the
magnetization will be calculated from Eq. \re{denk15}. Besides, many
of the thermodynamic functions can be obtained by solving Eqs.
\re{denk42} or  \re{denk44}. For instance, static hysteresis loops
can be obtained by obtaining the magnetization for different
magnetic field values ($H$) and the characteristics of them can be
determined such as hysteresis loop area, coercive field or remanent
magnetization. In addition, magnetic susceptibility of the system can be
obtained by numerical differentiation of the magnetization with respect to
the magnetic field.

Calculation with MFA-$N$ is rather clear but we need more elaboration
on the calculation with Eq. \re{denk44}. Eq. \re{denk44} contains
number of $N$ summations which run on the array of the dummy indices
$\{t_k\}\rightarrow \paran{t_1,t_2,\ldots, t_N}$. The dummy index of
$t_k$ takes the values of $-\delta_k,-\delta_k+2,\ldots,
\delta_k-2,\delta_k$, i.e. number of $\delta_k+1$ different values.
Thus,  Eq. \re{denk44} contains  number of
$\prodd{k=1}{N}{}\paran{\delta_k+1}$ terms to be summed.   Remembering
that, $\delta_k$ was the number of outer nearest neighbor spins of
the spin labeled by $S_k$. Any term in summation in Eq. \re{denk44},
has two parts which are being producted. First part is product of the
coefficients  $C_{t_k}$ which can be calculated from Eq.
\re{denk31}. The other part is the function evaluated at an ordered
array $\{t_k\}$ and this part can be calculated from Eq.
\re{denk12}. But in order to make calculations for any cluster, the
crucial point is to construct the configurations of the evaluation
points of the function, i.e. constructing the set of
$\paran{t_1,t_2,\ldots, t_N}$ from all possible values of any $t_k$.
The configuration set will have the number of
$\prodd{k=1}{N}{}\paran{\delta_k+1}$ different configurations of
ordered array $\{t_k\} $.

Similar strategy is valid for the calculation of Eq. \re{denk42}.
But it can be seen from Eqs. \re{denk42} and \re{denk43} that, the
number of configurations in which the function is evaluated is higher than
the procedure of calculation with Eq. \re{denk44}. As explained in
Sec. \ref{formulation1}, it will be better to use Eq. \re{denk44}
instead of Eq. \re{denk42} for the time saving during the numerical
processes.

For  obtaining the critical temperature of the system within the
EFT-$N$ or MFT-$N$ formulations given by Eqs. \re{denk42} or
\re{denk44} and Eq. \re{denk15}, respectively, linearized (in $m$)
forms of that expressions have to be obtained. Since in the vicinity
of the (second order) critical point, magnetization is very small,
the solutions of the linearized  equations for the temperature with
nonzero magnetization will give the critical temperature. As usual,
let us take into account the expression of the magnetization in a
form \eq{denk45}{ m=\summ{n=0}{N}{A_nm^n} }  then the linearized
form of Eq. \re{denk45} i given by \eq{denk46}{
\paran{1-A_1}m=0.
} Note that due to the time reversal symmetry of the system (i.e.
$H=0$ in Eq. \re{denk1})  $A_0=0$ has to be satisfied. The
temperature found from the solution of Eq. \re{denk46} (i.e. the
solution of $A_1=1$)  is critical temperature of the system.
Then it is important to obtain the coefficient $A_1$ for the
$N$-spin cluster from Eqs. \re{denk42} or \re{denk44}, in order to
get the critical temperature of the system within the EFT-$N$
formulation.  It is also important to get this coefficient for the
calculation within the EFRG, since the critical temperature can be
obtained by equating the coefficients $A_1$ with two different sized
clusters \cite{ref22}.

From the linearized form of Eq. \re{denk44}, the coefficient
$A_1$ can be obtained as

\eq{denk47}{ A_1^{EFT-N}=
\paran{\frac{1}{2}}^{\Delta}\summ{t_1=-\delta_1}{\prime
\delta_1}{}\summ{t_2=-\delta_2}{\prime \delta_2}{}\ldots
\summ{t_N=-\delta_N}{\prime \delta_N}{}
\left[\prodd{k=1}{N}{}\komb{\delta_k}{(\delta_k-t_k)/2}\right]\tau
F\paran{\{t_k J\}} } where \eq{denk48}{
\Delta=\summ{l=1}{N}{\delta_l} , \quad \tau=\summ{l=1}{N}{t_l} .}

On the other hand, linearization of Eq. \re{denk15} will give $A_1$ for the MFT-$N$ approximation as
\eq{denk49}{
A_1^{MFT-N}=\left. \ktur{F\paran{\beta,J,\{\delta_i J m\}}}{m}\right|_{m=0}
.}

\section{Results and Discussion}\label{results}

In this section, we want to present the effect of the working with
larger clusters on the critical temperatures and  some thermodynamic
properties of different lattices. For this aim we work on the two of
the two dimensional lattices, namely honeycomb and square lattices
and as an example of the three dimensional lattice, simple cubic
lattice. All these lattices have S-1/2 spins on their sites. Let us
define scaled temperature as $t=k_BT/J$ and  scaled critical
temperature as  $t_c=k_BT_c/J$, where $T_c$ is the critical
temperature.  Critical temperature within the EFT-$N$ formulation
can be obtained from the numerical solution of $A_1^{EFT-N}=1$ and
within the MFT-$N$ formulation from $A_1^{MFT-N}=1$, where
$A_1^{EFT-N}$ and $A_1^{MFT-N}$ are defined by Eqs. \re{denk47} and
\re{denk49}, respectively. On the other hand, within the MFRG
\cite{ref20}   and EFRG methods \cite{ref22}, critical temperatures
can be obtained from equations $A_1^{MFT-N}=A_1^{MFT-N^\prime}$
and $A_1^{EFT-N}=A_1^{EFT-N^\prime}$ for different cluster sizes
(number of spins which are inside and on the border in constructed
cluster) $N$ and $N^\prime$, respectively.

\subsection{Critical Temperatures}\label{results1}

In Fig. \re{sek1} we can see (a) the geometry of the honeycomb
lattice and (b) the variation of the critical temperature of the two
dimensional honeycomb lattice with the cluster size. Here $N$-spin
cluster has been constructed with the spins numbered from $1$ to $N$
in Fig. \re{sek1} (a). Firstly, we can see from Fig. \re{sek1}
(b) that, enlarging the cluster gives lower critical temperatures. At
the same time, lower values of the critical temperatures mean that,
more closer critical temperatures to the exact results. For this
lattice, the cluster size of $N=12$ in the EFT-$N$ formulation gives
the results of the BPA. Although the enlarging cluster lowers the
critical temperatures, this decreasing behavior of the critical
temperature when the size of the cluster rises, is not monotonic.
The same situation can be seen in Figs. \re{sek2} (b) and \re{sek3}
(b) for the square and the simple cubic lattices, respectively. The
cluster sizes of the square and simple cubic lattices which can give
the results of BPA within the EFT-$N$ formulation are $N=6$ and
$N=13$ respectively. Of course, when the coordination number of the
lattice rises, numerical calculations  of EFT-$N$ for larger
clusters becomes harder. This comes from the rising number of
evaluation points of the function given in Eq. \re{denk47}. These
numbers can be seen in Table \ref{tbl1}.

In order to investigate the behavior of the critical temperature
with the cluster size ($N$), we have fitted the critical
temperatures to the sizes of the cluster. It seems that the function
$t_c(N)=aN^{-b}$ is suitable form to mimic this behavior seen in
Figs. \re{sek1}-\re{sek3} (b). Here, $a=t_c(1)$ means that the one
spin cluster result for the critical temperature with the method
related to the curve, i.e. for the MFT, $a=3.0,4.0,6.0$ while for
the EFT $a=2.104, 3.090, 5.073$ \cite{ref12} for the  honeycomb,
square and simple cubic lattices, respectively. After the fitting
procedure we can find answers to the questions such as, how large
cluster is enough for obtaining the results of the BPA, which
cluster size gives the result that infinitely close to the exact
result?  Of course both of the methods cannot give the exact results
even if the cluster is really large, but finite. But, obtaining the
answer of the second question will give hints about the accuracy of
the results when the cluster size rises.


Fitting results of results of the both of the approximations
(MFT-$N$ and EFT-$N$) can be seen in Tables 3.2 and 3.3,
respectively. According to this fitting procedure, size of the
cluster that gives the results of the BPA ($N_{BPA}$) and results
that infinitely close to the exact result ($N_{exact}$) also given
in tables. It is not surprising to see that, EFT-$N$ reaches more quickly to
the results of BPA than the MFT-$N$, while enlarging the cluster.
For instance for the square lattice,  MFT-$37$ gives the BPA result
while in case of EFT, EFT-$6$ gives that result. But the interesting
point is in the values of $N_{exact}$. The values of the
$N_{exact}$ of the MFT are lower than that of the EFT, for all
lattices. It can be seen in fitting results in $b$ values in Tables
3.2 and 3.3 that the critical temperature values of the  MFT-$N$
decreases more quickly than the results of the EFT-$N$. But since
the MFT-$N$ curves starts with higher values than the EFT-$N$ curves
(i.e. the values of the $a$ parameters of the MFT-$N$ is higher than
the EFT-$N$), EFT-$N$ curves reach more quickly to the level of BPA.
But the higher rate of decrease of the curves MFT-$N$ results to
reaching the infinitely close to the exact results before the curves
of EFT-$N$.

\begin{figure}[h]\begin{center}
\epsfig{file=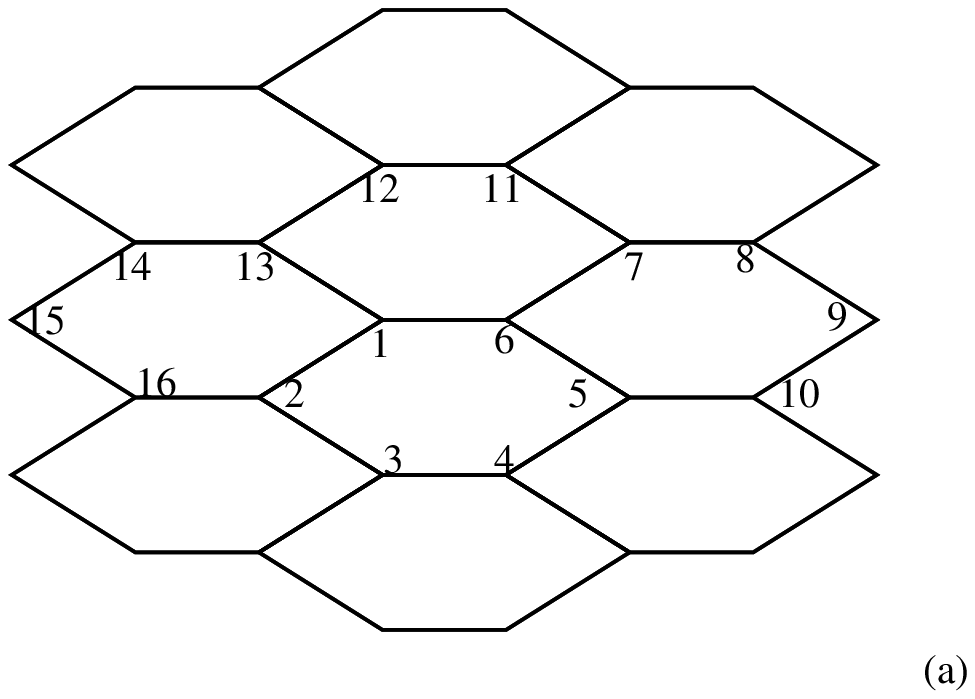, width=4.5cm}
\epsfig{file=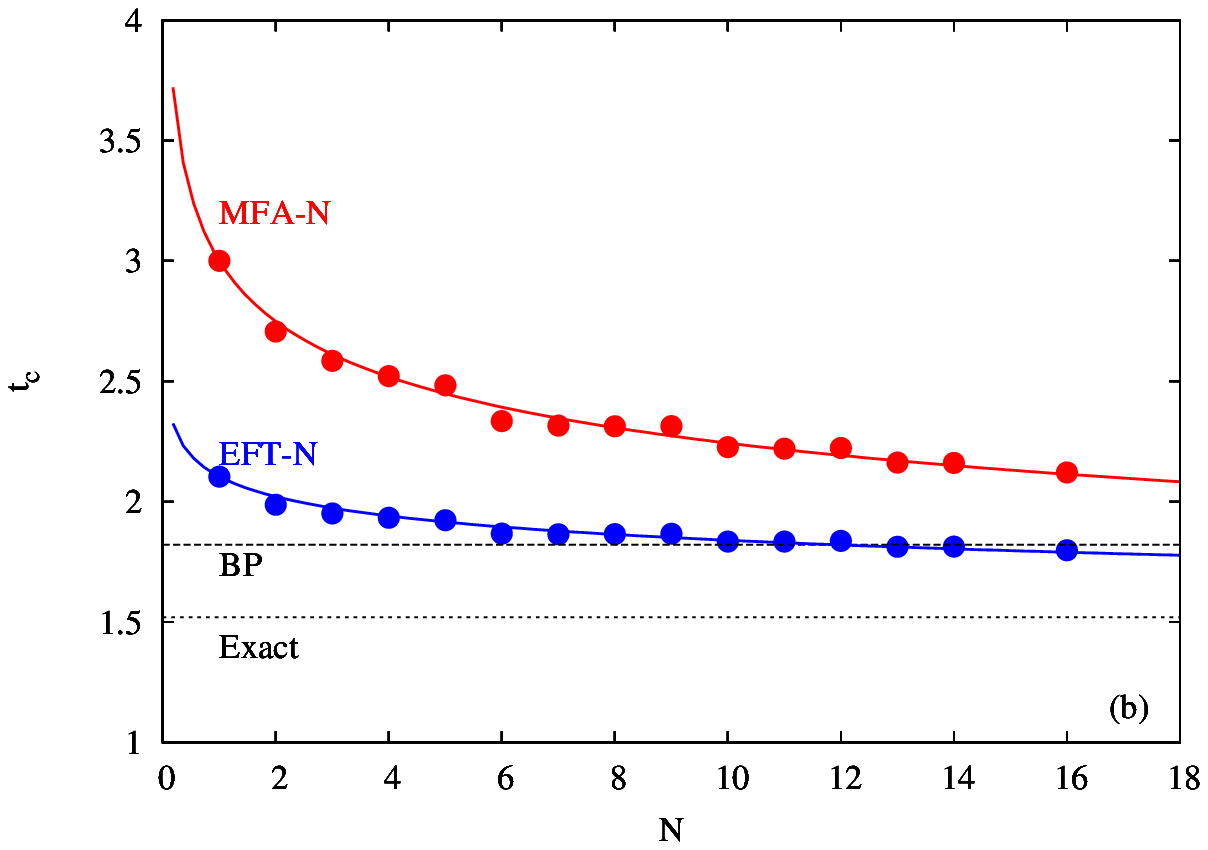, width=7.5cm}
\end{center}
\caption{(a) Schematic representation of the honeycomb lattice  (b)
the variation of the critical temperature with the size of the
cluster, for the honeycomb lattice. Exact result ($1.519$)
\cite{ref25} and the result of the BPA ($1.821$) \cite{ref26} also
shown as horizontal lines. Results are shown by points and also
fitted curve of the form $aN^{-b}$ depicted both of the methods
MFA-$N$ and EFT-$N$.} \label{sek1}\end{figure}

\begin{table}\label{tbl1}\caption{Number of elements in configuration set $\{t_k\}$
for Eq. \re{denk47}, for the honeycomb ($z=3$), square ($z=4$) and  simple cubic ($z=6$) lattices. }\ortala{
\begin{tabular}{||c|c|c|c||}
$N$&$z=3$&$z=4$&$z=6$\\
\hline
$1$&$4$&$5$&$7$\\
\hline
$2$&$9$&$16$&$36$\\
\hline
$3$&$18$&$48$&$180$\\
\hline
$4$&$36$&$81$&$625$\\
\hline
$5$&$72$&$216$&$3000$\\
\hline
$6$&$64$&$324$&$10000$\\
\hline
$7$&$96$&$864$&$32000$\\
\hline
$8$&$192$&$972$&$65536$\\
\hline

\end{tabular}}
\end{table}

\begin{figure}[h]\begin{center}
\epsfig{file=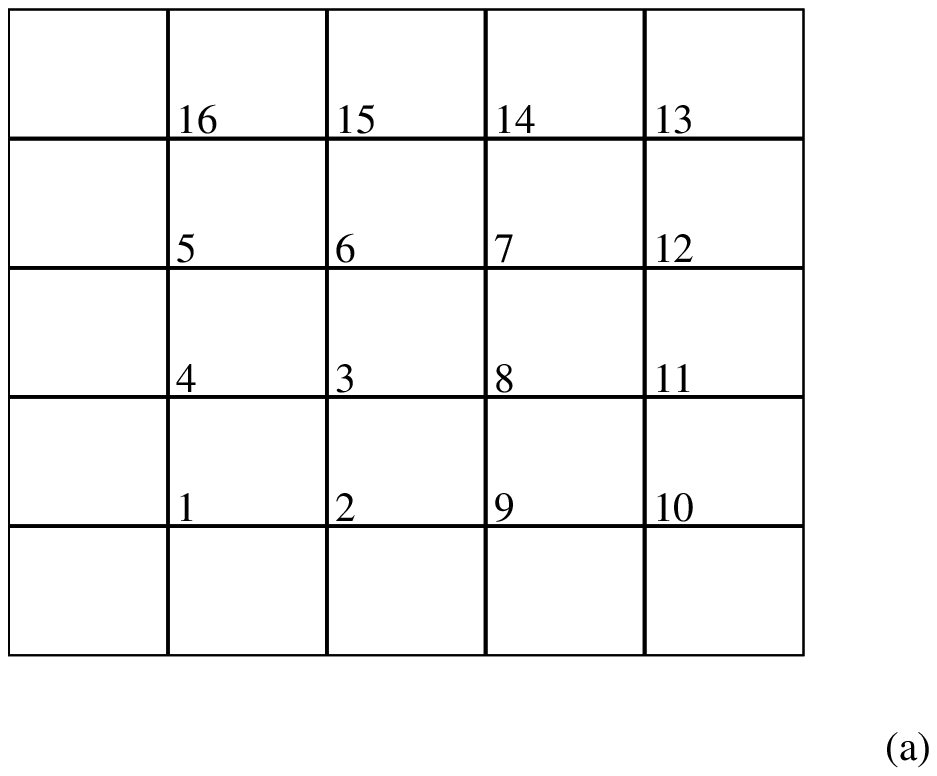, width=4.5cm}
\epsfig{file=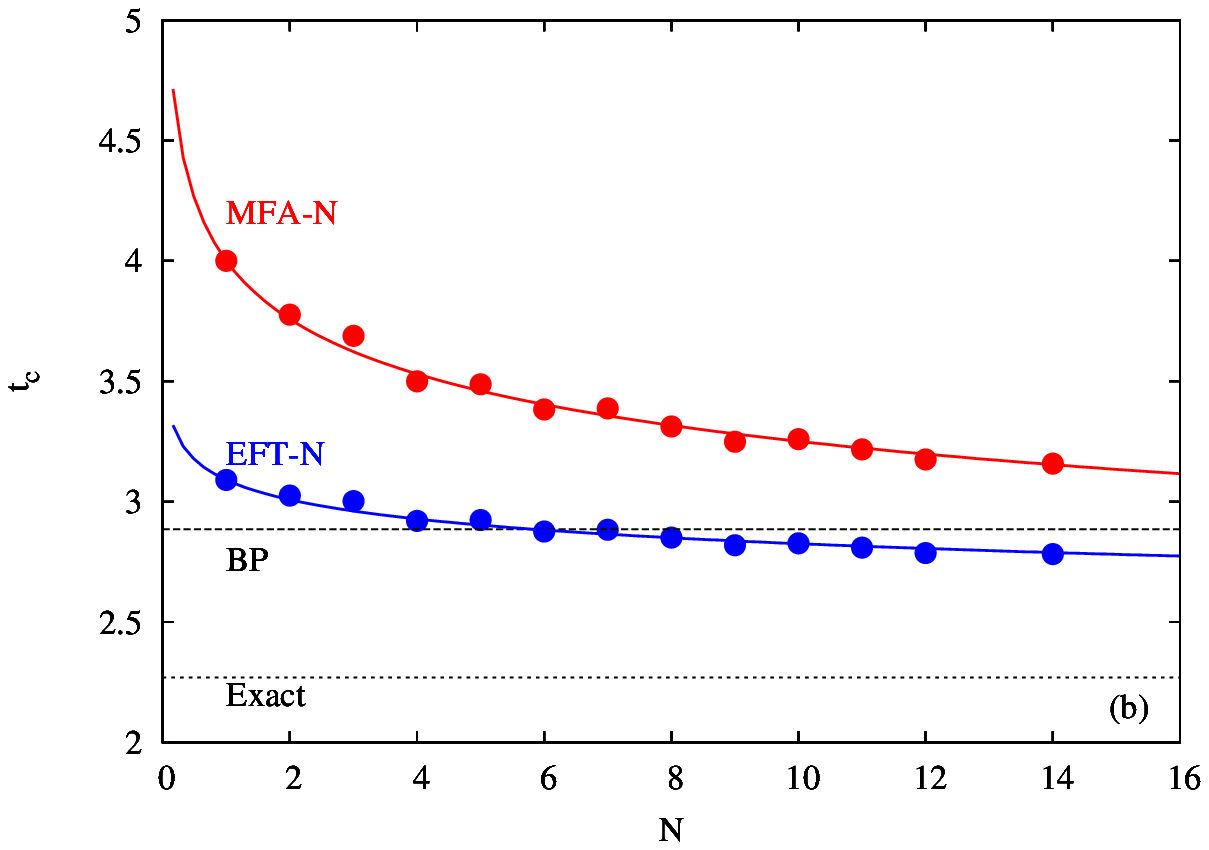, width=7.5cm}
\end{center}
\caption{(a) Schematic representation of the square lattice (b) the
variation of the critical temperature with the size of the cluster,
for the square lattice. Exact result ($2.269$) \cite{ref27} and the
result of the BPA ($2.885$) \cite{ref26} also shown as horizontal
lines. Results are shown by points and also fitted curve of the form
$aN^{-b}$ depicted for both of the methods MFA-$N$ and EFT-$N$.}
\label{sek2}\end{figure}

\begin{figure}[h]\begin{center}
\epsfig{file=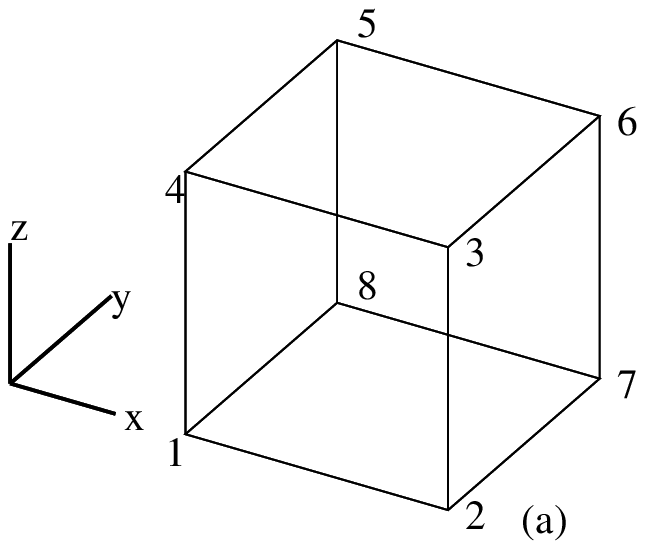, width=4.5cm}
\epsfig{file=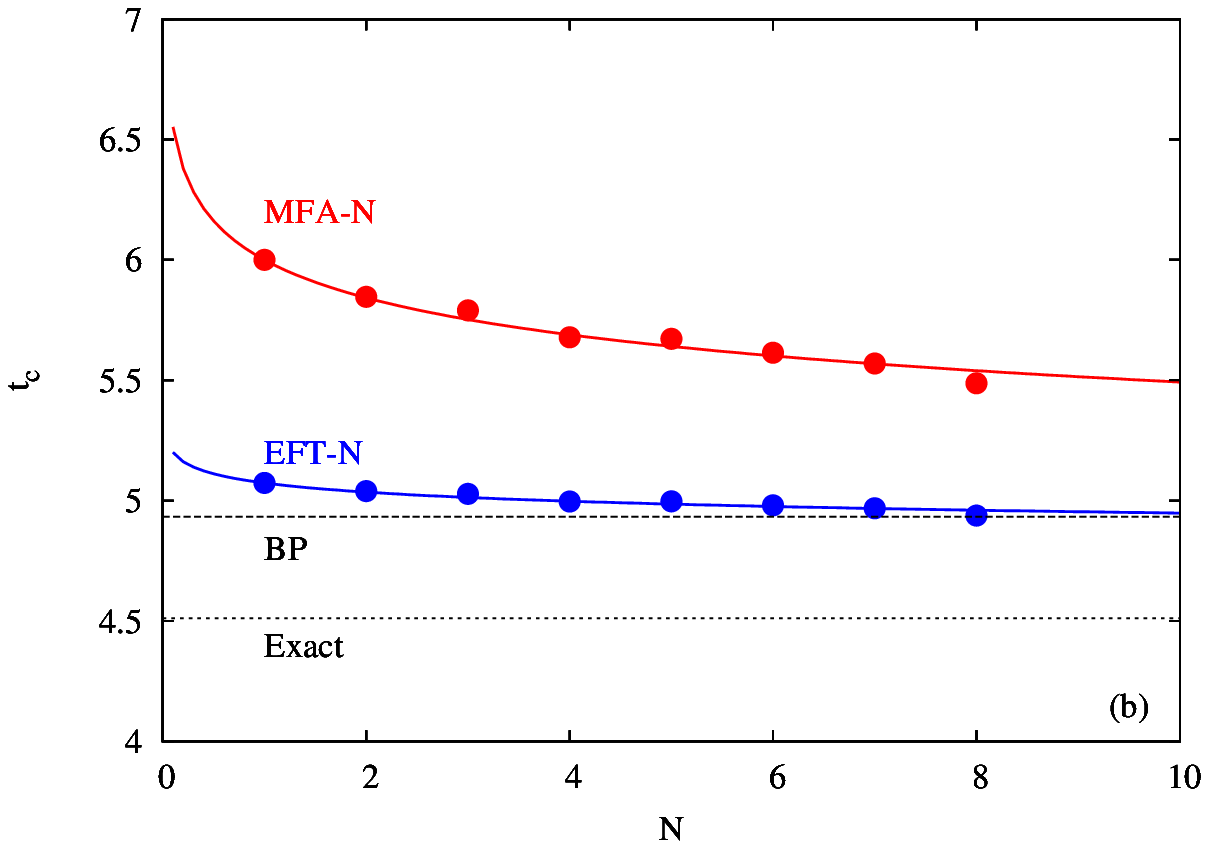, width=7.5cm}
\end{center}
\caption{(a) Schematic representation of the simple cubic lattice
(b) the variation of the critical temperature with the size of the
cluster, for the simple cubic lattice. Exact result ($4.511$)
\cite{ref25} and the result of the BPA ($4.933$) \cite{ref26} also
shown as horizontal lines. Results are shown by points and also
fitted curve of the form $aN^{-b}$ depicted both of the methods
MFA-$N$ and EFT-$N$.} \label{sek3}\end{figure}

\begin{table}\label{tbl2}\caption{$t_c=aN^{-b}$ least squares fitting results for the MFT-$N$ formulation}\ortala{
\begin{tabular}{||c|c|c|c|c||}
Lattice ($z$) &$b$&Sum of squares of residuals&$N_{BP}$&$N_{exact}$\\
\hline
$3$&$0.1262 $&$  0.0110 $&$52$&$217$\\
\hline
$4$&$ 0.0901$&$   0.0095 $&$37$&$537$\\
\hline
$6$&$0.0384$&$   0.0055$&$164$&$1691$\\
\hline

\end{tabular}}
\end{table}

\begin{table}\label{tbl3}\caption{$t_c=aN^{-b}$ least squares fitting results for the EFT-$N$ formulation}\ortala{
\begin{tabular}{||c|c|c|c|c||}
Lattice ($z$) &$b$&Sum of Squares of residuals&$N_{BP}$&$N_{exact}$\\
\hline
$3$&$ 0.0584 $&$ 0.0034 $&$12$&$259$\\
\hline
$4$&$0.0389$&$0.0036$&$6$&$2868$\\
\hline
$6$&$0.0108$&$  0.0009 $&$13$&$51478$\\
\hline

\end{tabular}}
\end{table}

As explained above, enlarging the cluster yields more accurate
results for the critical temperatures. But on some problems we have
to use larger clusters, even though we do not need the more accurate
results. Both of the approximations in $1$-spin cluster cannot
distinguish of some different lattice types. Most trivial example is
EFT-$1$ formulation can not distinguish the simple cubic lattice
from the triangular lattice, since both of the lattices have
coordination number (number of nearest neighbors) 6 and EFT-$1$ uses
only the coordination numbers. This deficiency may yield some
dramatic results. In order to explain this point, suppose that we
have a magnetic system with a  geometry given as Fig. \re{sek4}.
System is infinitely long about the $z$ axis and finite in $xy$
plane. With this geometry we can model the single walled nanotube.
In this form there are number of 6 spins in each plane. Beside the
present interaction between this nearest neighbor spins in one
plane, also there are interactions with  
nearest neighbor spins in the lower and upper planes. 
Let us call $L=6$ as the size of the nanotube,
which is the number of spins in each $xy$ plane. While in Fig.
\re{sek4} the size of the nanotube is 6, there can exist bigger or
lower sizes. For instance $L=3$ is a three-leg spin tube
\cite{ref28}.  Regardless of the size of the nanotube, if we solve
this system with EFT-$1$,  we obtain the results of the square
lattice. Because of  Eq. \re{denk26} (or Eq. \re{denk30}) contains
only the coordination number as a representation of the geometry of
the system. Then we have to enlarge the cluster. One of the
reasonable choice is to construct a finite cluster from the $L$ spins,
which are in the same plane. We can see the results for the critical
temperatures for this system in Fig. \re{sek5}. Constructed cluster
sizes and the size of the nanotube are the same, i.e. results taken
from the $L$-spin cluster, where the cluster consists of the spins
that belong to the one plane of the system. MFA-$1$ and EFT-$1$
results have been shown by horizontal lines in Fig. \re{sek5} with
the values $t_c=4.000$ and , $t_c=3.090$ respectively. As seen in
Fig. \re{sek5}, critical temperature rises when the size of the
nanotube gets bigger, as physically expected. But as seen in Fig.
\re{sek5}, 1-spin cluster formulations cannot give this situation.

\begin{figure}[h]\begin{center}
\epsfig{file=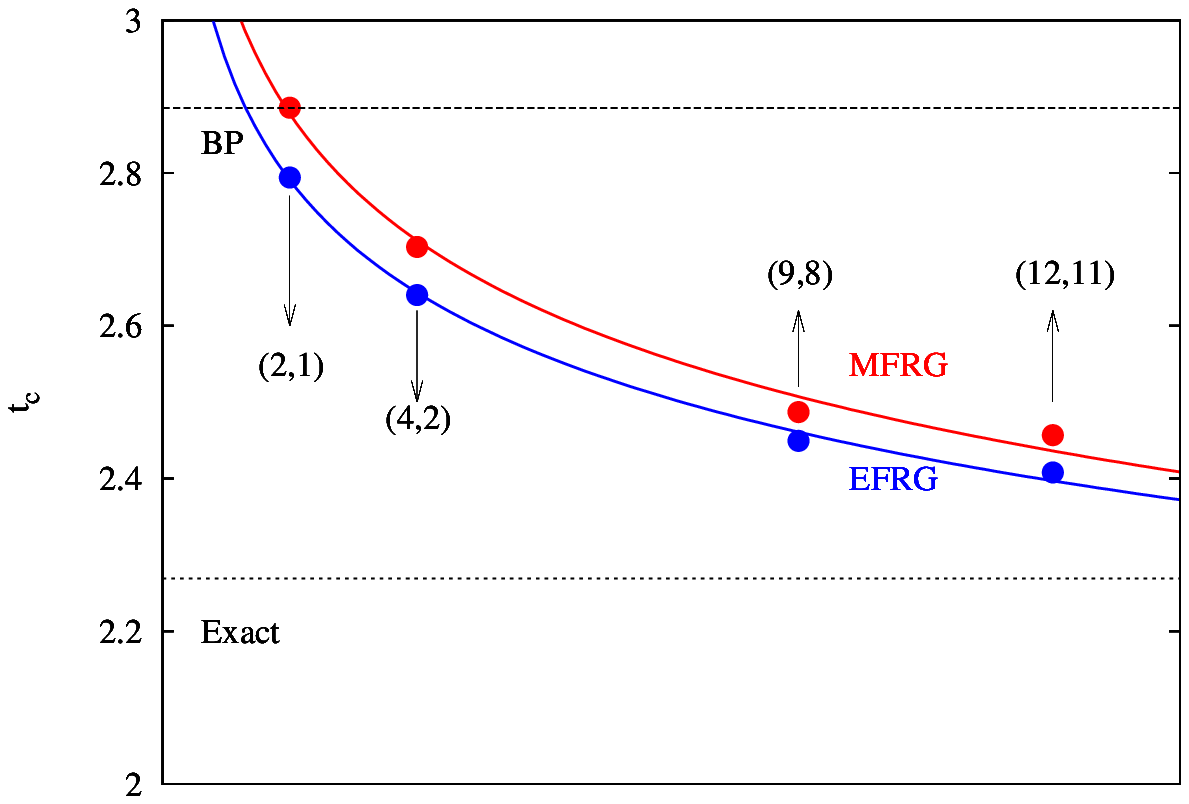, width=10cm}
\end{center}
\caption{Critical temperatures of the S-1/2 Ising model on square
lattice obtained from EFRG and MFRG methods. Cluster sizes that used
in both methods shown in parenthesis. The horizontal lines named as
BP and Exact are the results of BPA and exact calculations.}
\label{sek6}\end{figure}

Lastly, EFRG calculations on S-1/2 Ising systems can be easily done
by  using Eq. \re{denk47}. As an example of this, we have depicted
the variation of the critical temperature of the square lattice
(obtained within the EFRG formulation) with some selected cluster
sizes in Fig. \ref{sek6}.  As seen in Fig. \ref{sek6} that, critical
temperatures obtained from both of the methods (namely, EFRG and
MFRG) approach to the exact result,  while the size of the clusters
rise. Results for MFRG-$(2,1)$ ($t_c=2.885$), EFRG-$(2,1)$
($t_c=2.794$) are the same as given in Refs. \cite{ref29} and
\cite{ref22}, respectively.  On the other hand, the results of
EFRG-$(9,8)$ ($t_c=2.450$) and EFRG-$(12,11)$ ($t_c=2.408$) are
lower than the obtained value of EFRG-$(9,6)$ ($t_c=2.572$) in Ref.
\cite{ref23}. To the best of our knowledge, these last two results
have not been obtained within the EFRG yet.


\begin{figure}[h]\begin{center}
\epsfig{file=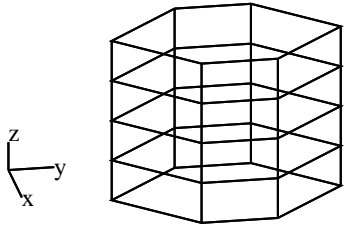, width=10cm}
\end{center}
\caption{Schematic representation of the single-walled nanotube with
size $L=6$.} \label{sek4}\end{figure}

\begin{figure}[h]\begin{center}
\epsfig{file=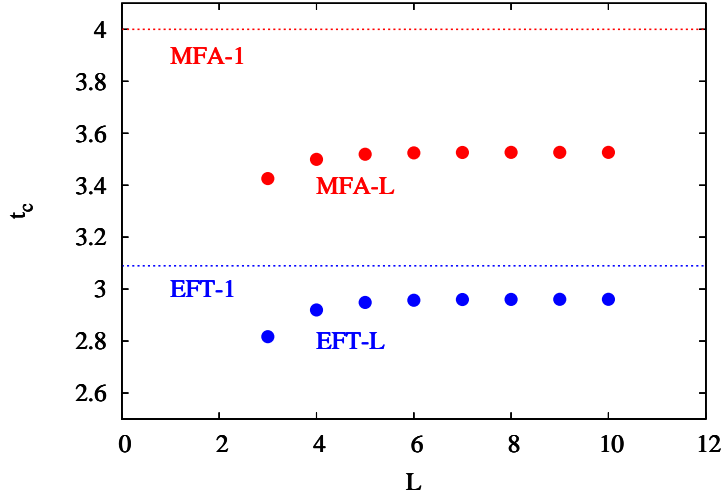, width=10cm}
\end{center}
\caption{Variation of the critical temperature of the single-walled
nanotube with the size of the nanotube, for both formulations
MFA-$L$ and EFT- $L$. The results of the  MFA-$1$ and EFT- $1$ also
shown with horizontal lines.} \label{sek5}\end{figure}

\subsection{Thermodynamic Properties}\label{results2}

In this section we want to investigate the effect of the enlarging
the cluster on the thermodynamic properties of the system. Since
different lattices have similar behaviors then we restrict ourselves
in only square lattice.

Magnetization can be calculated from Eq. \re{denk44} as explained in
Sec. \ref{formulation}. The differentiation of  Eq. \re{denk44} with
respect to magnetic field will give the magnetic susceptibility
($\chi$) of the system.  Besides, internal energy of the system
(denoted as $u$, which is scaled by $J$) can be calculated as the same
way of magnetization. The only difference is  the starting point of
the calculation i.e. in Eq. \re{denk8}, instead of $S_k$ there will
be terms like $S_kS_l$ which are the nearest neighbors of the
chosen cluster. Again, differentiation of this expression with
respect to the temperature will give the specific heat (denoted by
$c$, which is again scaled by $J$).

\begin{figure}[h]\begin{center}
\epsfig{file=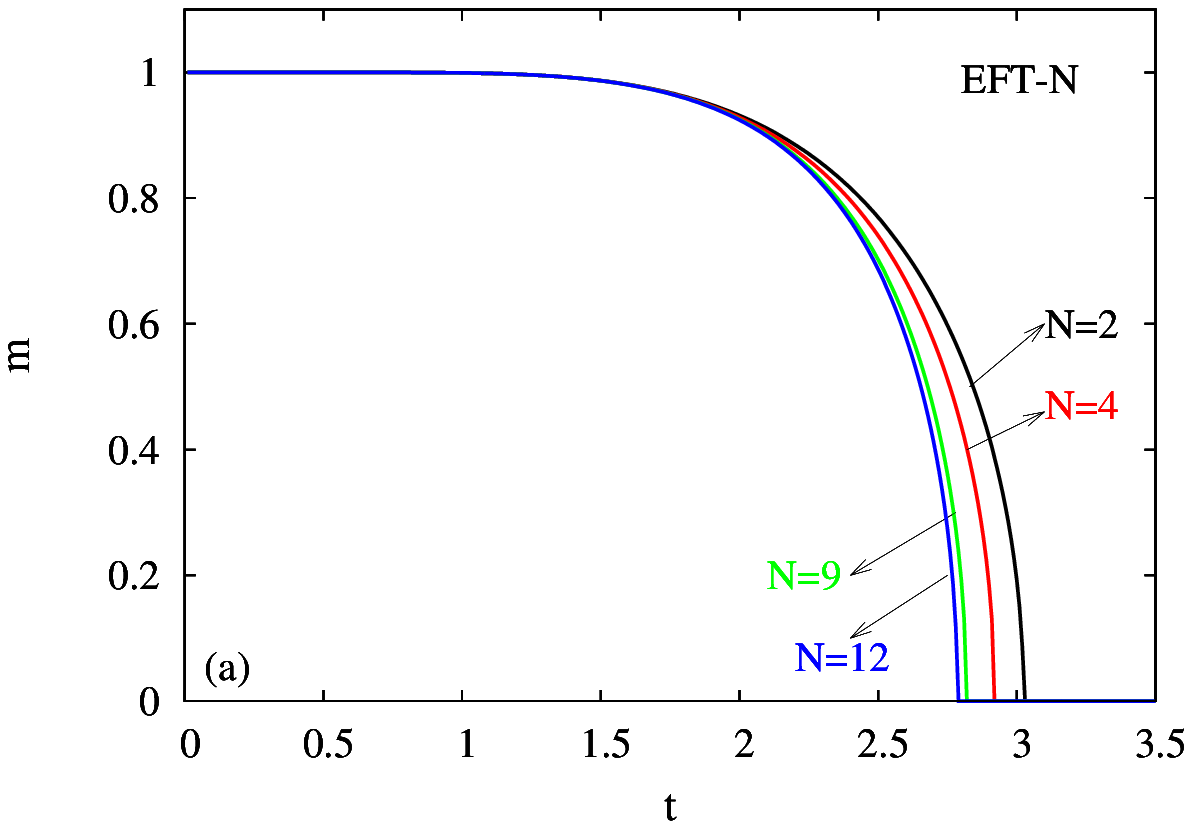, width=6cm}
\epsfig{file=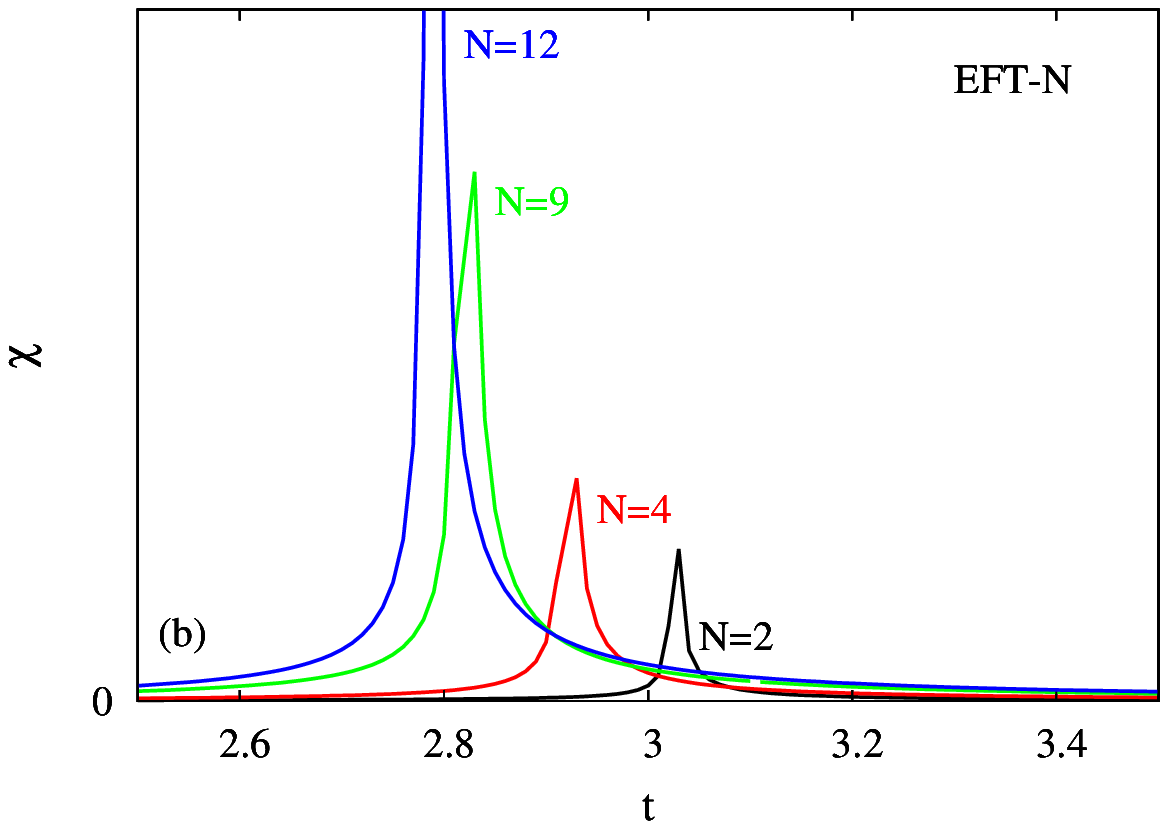, width=6cm}
\end{center}
\caption{Variation of the (a) zero-field magnetization and (b)
zero-field magnetic susceptibility  of the S-1/2 Ising model on a
square lattice, with the formulation EFT-$N$ and for selected values
of cluster sizes, $N=2,4,9,12$.} \label{sek7}\end{figure}

In order to see the effect of the enlarging cluster within the
EFT-$N$ formulation, we depict the variation of the magnetization
and the magnetic susceptibility of the system at zero magnetic
field, with the temperature for different cluster sizes in Fig.
\ref{sek7}. As seen in Fig. \ref{sek7} (a) the magnetization
behaviors with the temperature are the same for all of the clusters.
The only difference comes from the critical temperature, in which the
magnetization reaches to value of zero. As the size of the cluster
increases, the critical temperature decreases, as shown also in Fig.
\ref{sek2} (b). This decreasing behavior of the critical temperature
shows itself also in the  behavior of the magnetic susceptibility.
As seen in Fig. \ref{sek7} (b), while the size of the cluster
increases, the peaks of the susceptibility curves grow as well as
they shift to the right of the $(\chi-t)$ plane, i.e. lower
temperature regions. As we can see from \ref{sek7} (b) that,
enlarging cluster gives more realistic results for the magnetic
susceptibility, since the divergence behavior of the magnetic
susceptibility at a critical temperature appears more strongly  as
the size of the cluster rises.


\begin{figure}[h]\begin{center}
\epsfig{file=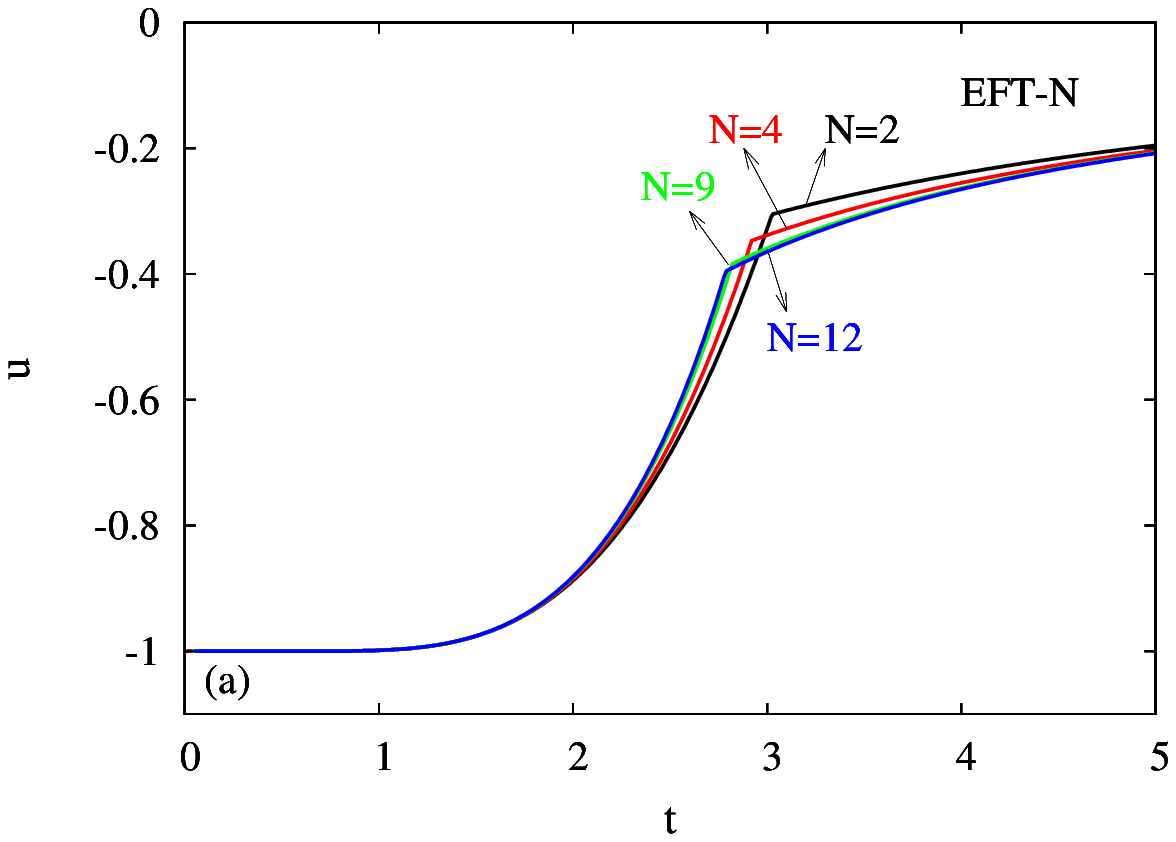, width=6cm}
\epsfig{file=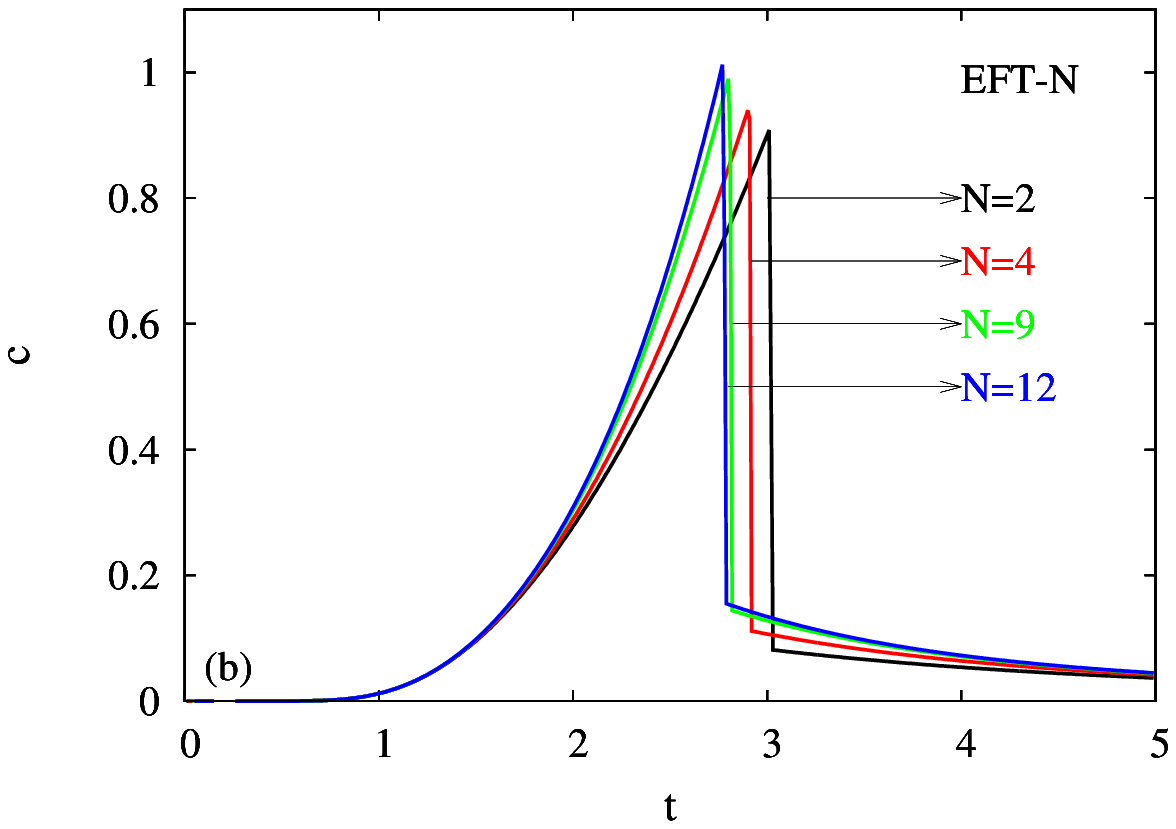, width=6cm}
\end{center}
\caption{Variation of the (a) internal energy and (b) specific heat
of the S-1/2 Ising model on a square lattice, with the formulation
EFT-$N$ and for selected values of cluster sizes, $N=2,4,9,12$.}
\label{sek8}\end{figure}

We can make similar conclusions about the behavior of the internal
energy and the specific heat of the system, when the size of the
cluster rises within the EFT-$N$ formulation. We can see from Fig.
\ref{sek8} (a) that, the change in the behavior of the internal
energy with temperature, occurs at lower values of the temperature
as the cluster size rises, since enlarging cluster causes to decline
of the critical temperature. The same thing shows itself in  Fig.
\ref{sek8} (b) also, which is the variation of the specific heat
with the temperature for some selected values of the cluster
sizes. The peaks, which occurs at the critical temperature, getting
higher when the cluster size rises.

All these comments suggest  that, within the EFT-$N$ formulation,
enlarging the cluster also will give more realistic results in the
thermodynamic properties of the system. As in the effect of the
enlarging cluster on the critical temperatures of the system, while
rising the size of the cluster, the difference between the
successive curves getting smaller.


\section{Conclusion}\label{conclusion}

In conclusion, a  general formulation for the EFT with differential
operator technique and DA (as well as MFT) with larger finite
clusters has been derived. Enlarging the finite cluster yields
different formulations which are called EFT-$N$ (or MFT-$N$) for the
$N$-spin cluster. The formulation is limited to the S-1/2 Ising
model on completely translationally invariant lattices.

It has been shown that, application of the EFT-$N$ and MFT-$N$
formulations on several lattices yield more accurate results in
critical temperatures as well as the thermodynamic properties of the
system, when the size of the cluster rises. Comparisons of the
results in the critical temperatures have been made with the results
of the BPA and exact ones. It has been shown that EFT-$6$ and
MFT-$37$ results  and EFT-$13$ and MFT-$164$ results in the critical
temperature, gives the results of the BPA for the square and simple
cubic lattices, respectively. We note here that, constructing
process of the finite cluster with $N$ spins can be made in several
ways. Different geometrical clusters which have the same number of
spins will give different results.

Besides, the limitations of the derived formulation have been discussed, since
enlarging the cluster yield more and more numerical computations, and
then takes more and more time. Anyway, we can say that the
formulation derived in this work can be applied to any cluster size, in
principle.

Besides all of these, derived formulation can be used in EFRG (and
MFRG) formulations. The effect of the enlarging cluster on the
critical temperatures of the square lattice within EFRG formulation
has been also discussed, with applying the formulation. The simplest
possible MFRG formulation gives the results of the BPA in the
critical temperature, while the EFRG results lie always below of the
MFRG results, as expected.

In addition to all of these observations, necessity of the using
$N$-spin cluster formulations in some systems (such as nano magnetic
systems) has been discussed. Constructing EFT-$N$ formulation for
the magnetic nano materials will be the topic of the future work.

We hope that the results  obtained in this work may be beneficial
form both theoretical and experimental point of view.


\newpage

\begin{thebibliography}{00}


\bibitem{ref1} R. J. Baxter, Exactly Solved Models in Statistical Mechanics (London: Academic Press, 1982).



\bibitem{ref2} E. Ising, Zeitschrift f\"{u}r Physik, 31,   (1925) 253.

\bibitem{ref3} L. Onsager, Phys. Rev. 65, 117 (1944)
\bibitem{ref4} J. S. Smart, Effective Field Theories of Magnetism, Saunders, London, 1966.
\bibitem{ref5} S. Mukhopadhyay, I. Chatterjee, Journal of Magnetism and Magnetic Materials 270 (2004) 247

\bibitem{ref6}  T. Oguchi, Progr. Theoret. Phys. (Kyoto) 13 (1955)
148.
\bibitem{ref7}  H. A. Bethe, Proc. Roy. Soc., London A 150 (1935) 552.

\bibitem{ref8}  R. F. Peierls, Proc. Canbridge Phil. Soc. 32 (1936) 477.

\bibitem{ref9} H. B. Callen, Phys. Lett. 4 (1963) 161.
\bibitem{ref10} M. Suzuki, Phys. Lett. 19 (1965) 267.





\bibitem{ref11}  R. Honmura, T. Kaneyoshi, J. Phys. C 12 (1979) 3979.

\bibitem{ref12}  T. Kaneyoshi, Acta Phys. Pol. A 83 (1993) 703.


\bibitem{ref13}  F. Zernike, Physica 7 (1940) 565.


\bibitem{ref14}  N. Matsudaira, J. Phys. Soc. Japan 35 (1973) 1593.
\bibitem{ref15}  N. Boccara, Phys. Lett. A 94 (1983) 185.



\bibitem{ref16} T. Balcerzak, J. Magn. Magn. Mater. 97, 152 (1991).

\bibitem{ref17} M. Saber, Chin. J. Phys. 35 (1997) 577. 

\bibitem{ref18} A. Bob\'{a}k, M. Ja\v{s}\v{c}ur
Phys. Stat. Sol. B 135,  (1986) K9.


\bibitem{ref19}  O. R. Salmon ,  J. R. de Sousa  and F. D.  Nobre   Physics Letters A  373 (2009) 2525


\bibitem{ref20} J.O. Indekeu, A. Maritan, A.C. Stella, J. Phys. A 15 (1982) L291.

\bibitem{ref21} V. Ilkovi\v{c}, Phys. Stat. Sol. (B), 166  (1991) K31.
\bibitem{ref22} I.P. Fittipaldi, D.F. de Albuquerque, J. Magn. Magn. Mater. 107 (1992) 236.
\bibitem{ref23} D.F. de Albuquerque, E. Santos-Silva, N.O. Moreno J. Magn. Magn. Mater. L63 (2009) 321.

\bibitem{ref24} \"{U}. Ak\i nc\i \   arXiv:1308.2511v2 (2014).


\bibitem{ref25}  M.E. Fisher, Rep. Prog. Phys. 30, (1967) 615.


\bibitem{ref26}   T. Kaneyoshi, Physica A 269, (1999) 344.


\bibitem{ref27}  L. Onsager, Phys. Rev. 65, (1944) 197.

\bibitem{ref28}  T. Sakai, M. Sato , K. Okamoto, K. Okunishi and C. Itoi J. Phys.: Condens. Matter 22 (2010) 403201


\bibitem{ref29} J.O. Indekeu, A. Martian, A.L. Stella, Phys. Rev. B 35 (1987) 305.




\end{thebibliography}
\end{document}